\documentclass{moriond}

\def\PRD{{\em Phys. Rev.} D}

\def\be{\begin{equation}}
\def\ee{\end{equation}}
\def\bea{\begin{eqnarray}}
\def\eea{\end{eqnarray}}

\newcommand{\Photo}{}

\begin{document}
\vspace*{4cm}
\title{ESPRESSO'S EARLY COMMISSIONING RESULTS AND PERFORMANCE RELATED TO TESTS OF FUNDAMENTAL CONSTANT STABILITY}

\author{A. C. O. LEITE$^{a,b}$, C. J. A. P. MARTINS$^{a}$, P. MOLARO$^{c}$, S. MONAI$^{c}$, C. S. ALVES$^{a,d}$, T. A. SILVA$^{a,b}$ and the ESPRESSO Science Team }

\address{$^{a}$Instituto de Astrof\'{i}sica e Ci\^{e}ncias do Espa\c{c}o, CAUP, Rua das Estrelas, 4150-762 Porto, Portugal\\ $^{b}$Faculdade de Ci\^{e}ncias da Universidade do Porto, Rua do Campo Alegre, 4150-007 Porto, Portugal\\ $^{c}$INAF-Osservatorio Astronomico di Trieste, Via G.B. Tiepolo 11, I-34143 Trieste, Italy\\
$^{d}$Department of Mathematics, Imperial College London, London SW7 2AZ, United Kingdom}

\maketitle\abstracts{ESPRESSO is a new high-resolution ultra-stable spectrograph for  the VLT, which had its first light on Telescope on November 27th, 2017.  The instrument is installed in the Combined Coud\'{e} Laboratory and linked to the 4 Units of Telescope through optical Coud\'{e} Trains, being the first spectrograph able to collect the light from the 4 UTs simultaneously.
 One of the key science goals of the instrument is to test the stability of nature's fundamental couplings with unprecedented resolution and stability.  ESPRESSO will allow to eliminate current known systematics and test the claim by Webb et al 2012 of a spatial dipole in the variation of the fine-structure constant. These improved results (either null or variation detections) will put strong constraints on a range of cosmological and particle physics parameters.
}

%\section{Guidelines}

%The Moriond proceedings are printed from camera-ready manuscripts.
%The following guidelines are intended to get a uniform rending of the 
%proceedings. Authors with no connection to \LaTeX{} should use this
%sample text as a guide for their presentation using their favorite
%text editor (see section~\ref{subsec:final})

\section{Introduction}

Quasar (QSO) absorption spectra are powerful laboratories to test the variation of fundamental constants. Absorption lines produced by the intervening clouds along the line of sight of the QSO give access to physical information on the atoms present in the cloud, and this means that they give access to physics at different cosmological times and places. Different energy levels of some atomic transitions in these clouds are sensitive to $\alpha$ variations and this sensitivity varies according to the different atomic complex structures. If the physics in the cloud where the absorption line is produced was different, then one would get position shifts compared with the laboratory wavelengths corrected for the redshift. We fit as many sensitive lines as possible to break degeneracies and to increase the statistical signal of the measurement.

The best measurements of $\alpha$ in QSO spectra existing today were obtained with two high-resolution spectrographs: UVES - Ultraviolet and Visual Echelle Spectrograph on the VLT - Very Large Telescope; HIRES - High Resolution Echelle Spectrometer, on the Keck observatory.
A data set of 293 archival measurements from these two instruments show a spatial variation of $\alpha$ at more than four-sigma level of statistical significance, at the ppm level.\cite{webb2011}  There is no identified systematic effect that is able to fully explain it. A detailed analysis of the data reduction, dipole computation and search for systematics is described in King 2011.\cite{king}

After  this study some systematics in the wavelength calibration of data collected with high-resolution, slit spectrographs  were identified.
Rahmani et al. (2013) found long-range wavelength distortions in the VLT/UVES instrument by comparing solar spectra as reflected off asteroids with a Fourier Transform Spectrometer (FTS) solar flux spectrum. Other following studies found these distortions and showed that they vary over time.\cite{bagdonaite2014,evans2014,songaila2014,whitmore2015} Using 20 years of UVES and HIRES archival data, Whitmore \& Murphy \cite{whitmore2015} reported that the long range distortions vary between $\pm 200 m s^{-1}$ per $1000$ \AA. These effects could imply significant errors within the previous data set of measurements, weakening the evidence for the claim of the dipole variations in $\alpha$.

\section{ESPRESSO Commissioning}

ESPRESSO is a high-resolution, ultra-stable spectrograph. Being fiber-fed will allow for precise $\alpha$ measurements with control of the previous identified systematics. The instrument is installed at the  VLT and it allows to observe with  one  Unit of Telescope (UT)  individually  or to combine the light of the four telescopes. The instrument is installed in the Combined Coud\'{e} Laboratory and linked to the 4 Units of Telescope through optical Coud\'{e} Trains, being the first spectrograph able to collect the light from the 4UTS simultaneously. First light on 1 UT of the VLT was on November 27th, 2017, and  the 4 UT combined was on February 3rd, 2018.  The instrument is presently under commissioning. The 1UT tests will finish in July 2018 and the 4UT will last till the end of 2018.

On the ESO Call for Proposals for Period 102 (March 1st, 2018) the instrument was offered to the community for regular operations in 1UT mode only, this will allow observations in both HR (High Resolution) and UHR (Ultra High Resolution) modes. The ESPRESSO 4UT mode may be offered in Period 103, depending on successful commissioning during Period 101. 
%The specifications of  ESPRESSO are presented in Table \ref{tb:espresso} in each one of the three modes of operation. More detailed information of the technical specifications and observing modes can be found on the ESPRESSO User Manual: VLT-MAN-ESP-13520-253, Issue 0.8 \footnote{\url{www.eso.org/sci/facilities/paranal/instruments/espresso/ESPRESSO_User_Manual_v0.8.pdf}}.
The specifications of  ESPRESSO in the different modes of operation and more detailed information of the technical specifications can be found on the ESPRESSO User Manual: VLT-MAN-ESP-13520-253, Issue 0.8 \footnote{\url{www.eso.org/sci/facilities/paranal/instruments/espresso/ESPRESSO_User_Manual_v0.8.pdf}}.

%
%\begin{table}[h]
%%\captionsetup{width=1\textwidth}
%\caption[ ]{ \small Summary of ESPRESSO's instrument modes and corresponding performance. \cite{espresso} }
%\label{tb:espresso}
%\centering
%\begin{small}
%\begin{tabular}{|l|c|c|c|}
%\hline
% 	Parameter/Mode & singleHR (1 UT) & multiMR (up to 4 UTs) & singleUHR (1 UT)\\
% 	\hline
%Wavelength range & $380-780$ nm & $380-780$ nm & $380-780$ nm \\
%Resolving power (median) & $140 000$ & $70 000$  & $190 000$ \\
%Aperture on sky & $1.0$ arcsec & $4 \times 1.0$ arcsec & $0.5$ arcsec \\
%Global efficiency @550nm& $9\%$ & $9\%$ & $4\%$ \\
%Instrumental RV precision & $< 10 cm s^{-1}$ & $< 5 cm s^{-1}$ & $< 5 cm s^{-1}$  \\
%Limiting V-band magnitude & ~17& ~20 & ~16 \\
%Binning &1$\times$1, 2$\times$1 & 1$\times$1& 4$\times$2, 8$\times$4\\
%Spectral sampling (average) & $4.5$ pixels & $5.5$ pixels (binned$ \times 2$) & $2.5$ pixels \\
%Spatial sampling per slice & $9.0$ ($4.5$) pixels & $5.5$ pixels (binned$ \times 4$) & $5.0$ pixels  \\ 
%\hline
%\end{tabular}
%\end{small}
%\end{table}

The first commissioning results report good absolute calibration and stability of the instrument. For the fundamental constants tests it was crucial  to verify the absence of  the long-range drifts observed in previous instruments. 

\begin{figure}[h]
\centerline{\includegraphics[width=0.65\linewidth]{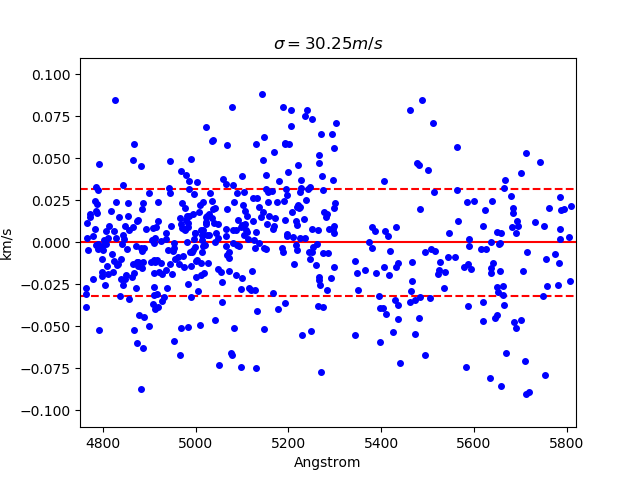}}%\vspace{-0.4cm}
\caption[]{\small Radial velocity differences between the solar lines as measured in the ESPRESSO 1UT Single 2x1 mode spectrum of Iris (6-Dec-2017) and the CERES-HARPS solar spectrum calibrated with the Laser Frequency Comb from \cite{molaro2013}. The gap at 530 nm is due to the HARPS gap.}
\label{fig:asteroid}
\end{figure}

Beside the long-range distortions, intra-order distortions have been detected in HARPS with the first Laser Frequency Comb (LFC) analysis\cite{wilken2010,molaro2013}.  The line comparison with the solar spectrum calibrated with the LFC (i.e. free from distortions), allow to check if they are present in ESPRESSO. The Figure 1 shows the differences for the 570 lines of Iris asteroid  6 Dec spectrum in the 4800-5800\AA ~range proved by the LFC. From the Figure 1 no long range slopes can be recognized. The Figure 2 shows orders 117 and 122 as examples showing no  evidence for  intra-order distortions in ESPRESSO spectra.

\begin{figure}[h]
\centerline{\includegraphics[width=1.15\linewidth]{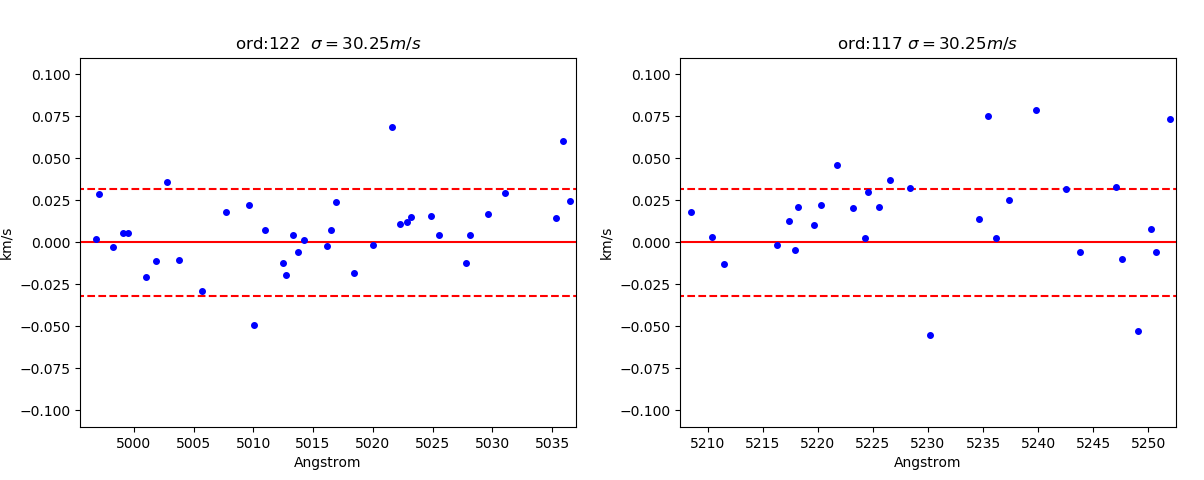}}%\vspace{-0.4cm}
\caption[]{\small Detailed radial velocity differences as in Figure 1 for orders 117 and 122 of the ESPRESSO spectrum.}
\label{fig:asteroid}
\end{figure}

This instrument has as one of its science goals to test the stability of fundamental couplings.\cite{espresso} The Fundamental Physics Guaranteed Time of Observation (GTO) for this purpose will have 27 nights and a list of 14 QSO "ideal" targets was put together.\cite{leiteGTO} In order to achieve stronger constrains on the variation of the fine-structure constant, $\alpha$,  an ideal quasar  target should present simple and strong absorption features of  transitions with high sensitivities  to variations of $\alpha$.
%
%\begin{table}[h]
%\caption[ ]{List of the best measurements of the stability of the fine structure constant considering the wavelength coverage of ESPRESSO. Column 1 gives the quasar name; the redshifts of the absorption system are given in Column 2; Column 3 gives the magnitude of the emitting quasar. Column 4 and 5 gives the value of the  measurement and the corresponding uncertainty.  The last Column gives the references for each measurement.}
%\label{tb:targetalpha}
%\vspace{0.1cm}
%\begin{small}
%\begin{center}
%\begin{tabular}{|c|c|c|c|c|c|}
%\hline 	
% 	Name & $z_{abs}$ & M & ${\Delta \alpha}/{\alpha}$ {\small $(10^{-6})$}  & $\sigma_{{\Delta \alpha}/{\alpha}} ${\small $(10^{-6})$} & Ref. \\
% 	\hline 		 
%J034943-381031 & 3.02 & 17.3& -27.9 & 34.2 & \cite{murphy} \\ 
%J040718-441013 &  2.59  & 17.3 & 5.7  & 3.4 & \cite{king} \\
%J043037-485523 &  1.35 & 16.5 & -4.0  & 2.3 & \cite{king}\\
% J053007-250329 & 2.14 & 18.8 &6.7  &3.5 & \cite{king}\\
% J110325-264515 & 1.84 & 15.9&5.6  &2.6  & \cite{lev2} \\
% J115944+011206 & 1.94 & 17.5  &5.1  &4.4  & \cite{king} \\
%J133335+164903 & 1.77  & 16.7 &8.4  &4.4 & \cite{king} \\
%HE1347-2457 	& 1.43 & 16.3 & -21.3 & 3.6   & \cite{lev2}\\
%J220852-194359 & 1.92 & 17.0 &8.5 & 3.8 & \cite{king} \\
%HE2217-2818 & 	1.69 & 16.0   & 1.3 & 2.4  & \cite{lp1}\\
%Q2230+0232 & 	1.86 & 18.0  & -9.9 & 4.9 & \cite{murphy} \\
%J233446-090812 & 2.15& 18.0 & 5.2 & 4.3 & \cite{king}\\
%J233446-090812 & 2.28 & 18.0 & 7.5 & 3.7   & \cite{king}\\
%Q2343+1232 & 	2.43  & 17.5  & -12.2 & 3.8 & \cite{murphy}\\
%\hline 
%\end{tabular}
%\end{center}
%\end{small}
%\end{table}

The 14 targets selected for the GTO regarding fine-structure constant together with the selection criteria  are presented in Leite et al. 2016\cite{leiteGTO} and the full list of protected targets for ESPRESSO can be seen online\footnote{ \url{www.eso.org/sci/observing/teles-alloc/gto/102.html}}. QSO B0347-383, one of the selected targets, is specially interesting since besides an $\alpha$ measurements it allows measurements of the proton-to-electron mass ratio and the  $T_{CMB}$ at the absorber redshift. This makes it an interesting target to investigate the nature of the fundamental constants variation testing different theories where a relation between these three constants is predicted.\cite{ferreira} The Consortium will observe these targets during the next 4 years and will aim to reach uncertainties bellow the ppm level, with already proven control of previous existing systematics. 

\section{Forecasts on Cosmology - Extension of Bekenstein-type models}
This type of measurements are important by themselves for testing the stability of $\alpha$ but even a non-detection of variation  can be used to constrain dark energy.

The Bekenstein-Sandvik-Barrow-Magueijo \cite{bekenstein} model assumes a variation of $\alpha$ that comes  from the variation of the electric charge. This  model can be seen as a $\Lambda$CDM-like model with an additional dynamical degree of freedom, whose dynamics is such that it leads to the aforementioned variations without having a significant impact on the Universe's dynamics. An extention of this type of model was discussed by Olive and Pospelov\cite{olive2002}, allowing for  different couplings to the dark matter and dark energy sectors and the behavior of $\alpha$ depends on both of them.
This would immediately suggest that the two parameters will be degenerate, but a combination of astrophysical measurements of $\alpha$, at approximate redshift $0 < z < 4$, and local laboratory tests partially breaks
this degeneracy and leads to strong constraints on both parameters. 
Constraints on these kinds of models from current astrophysical and cosmological data as well as detailed forecasts for ESPRESSO are presented in C. S. Alves et al. 2018\cite{alves2018}. The important variables are the couplings $\zeta_{m}$ and $\zeta_{\Lambda}$,  free parameters of the model that give magnitude to the allowed variation on $\alpha$.

Assuming the baseline and ideal uncertainties on $\alpha$ expected for  ESPRESSO, and applying these uncertainties for the 14 targets on the GTO target list The constrains on the two free parameters of this model are presented on Table \ref{tab3}. We compare it with the constrains of all the existing measurements of $\alpha$ ($\sim300$) and we can see that the  uncertainty retrieved from only 14 measurements of the GTO will be comparable with  the complete data set of already existing measurements.

\begin{table*}
\centering
\caption{\small\label{tab3}One sigma uncertainties on the couplings $\zeta_m$ and $\zeta_\Lambda$ (marginalizing the other) from existing $\alpha$ data, and the corresponding forecasts for the ESPRESSO Fundamental Physics GTO target list.  Uncertainties are presented in parts per million.}
\begin{small}
\begin{tabular}{|l|c c|c c|c c|}
\hline
Data set & $\delta\zeta_m$ & $\delta\zeta_\Lambda$ \\
\hline
Current constraints & $0.72$ & $1.84$ \\ 
\hline
ESPRESSO Baseline  & $0.73$ & $4.36$\\ 
ESPRESSO Ideal  & $0.24$ & $1.45$ \\
\hline
\end{tabular}
\end{small}
\end{table*}

\section*{Acknowledgments}
This work was done in the context of grant PTDC/FIS/111725/2009 from FCT (Portugal). ACL is supported by an FCT
fellowship (SFRH/BD/113746/2015), under the FCT PD Program PhD::SPACE (PD/00040/2012), and by the
Gulbenkian Foundation through Programa de Est\'{i}mulo  \`{a} Investiga\c{c}\~{a}o 2014, grant number 2148613525. CJM. is also supported by an FCT Research Professorship, contract reference IF/00064/2012, funded by FCT/MCTES (Portugal) and POPH/FSE (EC). 

\section*{References}

%\begin{thebibliography}{99}
%\bibitem{ja}C Jarlskog in {\em CP Violation}, ed. C Jarlskog
%(World Scientific, Singapore, 1988).
%
%\bibitem{ma}L. Maiani, \Journal{\PLB}{62}{183}{1976}.
%
%\bibitem{bu}J.D. Bjorken and I. Dunietz, \Journal{\PRD}{36}{2109}{1987}.
%
%\bibitem{bd}C.D. Buchanan {\it et al}, \Journal{\PRD}{45}{4088}{1992}.
%
%\end{thebibliography}

\end{document}